\documentclass[sigconf,10pt]{acmart}

\renewcommand\footnotetextcopyrightpermission[1]{} 
\usepackage{epsfig,endnotes}
\usepackage{csvsimple}
\usepackage{color} 
\usepackage{xcolor}
\usepackage{microtype}
\usepackage{vwcol}
\usepackage{multirow}
\usepackage{multicol}
\usepackage{pifont}
\usepackage{caption}
\usepackage{subcaption}
\usepackage{graphicx}
\usepackage{listings}
\usepackage{array}
\usepackage{mdwmath}
\usepackage{mdwtab}
\usepackage{float}
\usepackage{morefloats}
\newfloat{program}{t}{lop}
\usepackage{fixltx2e}
\usepackage{xcolor}
\usepackage{xspace}
\usepackage{balance}

\usepackage{url}
\usepackage{breakurl}
\expandafter\def\expandafter\UrlBreaks\expandafter{\UrlBreaks\do-\do_}
\usepackage{makecell}

\usepackage[utf8]{inputenc}

\newcommand{\myparagraph}[1]{\smallskip \noindent{\bf {#1}.}}

\newcommand{\out}[1] {}

\newcounter{codeLineCntr}

\reversemarginpar
\newif\ifnotes
\notestrue

\newcommand{\punt}[1]{}

\renewcommand{\eqref}[1]{Equation~(\ref{eq:#1})}

\newcommand{\proc}[1]{\ifmmode\mbox{\textsc{#1}}\else\textsc{#1}\fi}

  \newcommand{\func}[1]{\ifmmode\mathrm{#1}\else\textrm{#1}fi} %

\newcounter{remark}[section]

\usepackage[small,compact,noindentafter]{titlesec} 

\titlespacing\section{0pt}{6pt plus 2pt minus 2pt}{2pt plus 2pt minus 2pt}
\titlespacing\subsection{0pt}{4pt plus 2pt minus 2pt}{2pt plus 1pt minus 1pt}
\titlespacing{\paragraph}{0pt}{2pt plus 0pt minus 1pt}{1.0ex}

\usepackage[inline]{enumitem}
\setlist{noitemsep,topsep=0pt,parsep=0pt,partopsep=0pt}

\usepackage[subtle]{savetrees}

\usepackage{setspace}
\usepackage[margin=5pt,font={stretch=0.9}]{caption}

\newcommand{\sys}{\textsc{Slick}\xspace}
\newcommand{\slick}{\textsc{Slick}\xspace}
\newcommand{\scone}{\textsc{Scone}\xspace}
\newcommand{\click}{\textsc{Click}\xspace}
\newcommand{\sgx}{\textsc{SGX}\xspace}
\newcommand{\dpdk}{\textsc{DPDK}\xspace}
\newcommand{\intel}{\textsc{Intel}\xspace}

\begin{document}
\title{Slick: Secure Middleboxes using Shielded Execution}

\begin{abstract}
Cloud computing offers the economies of scale for computational resources with the ease of management, elasticity, and fault tolerance. To take advantage of these benefits, many enterprises are contemplating  to outsource the middlebox processing services in the cloud.   However,  middleboxes that process confidential and private data cannot be securely deployed in the untrusted environment of the (edge) cloud.  

To securely outsource middleboxes to the cloud, the state-of-the-art systems advocate network processing over the encrypted traffic.  Unfortunately, these systems support only restrictive middlebox functionalities, and incur prohibitively high overheads due to the complex computations involved over the encrypted traffic.

This motivated the design of \sys{}---a secure middlebox framework for deploying high-performance Network Functions (NFs) on untrusted commodity servers.  \sys exposes a generic interface based on \click to design and implement a wide-range of NFs  using its out-of-the box elements and C++ extensions.  \sys leverages \scone (a shielded execution framework based on \intel \sgx) and \intel \dpdk to securely process confidential data at line rate. 

More specifically, \slick provides hardware-assisted memory protection, and configuration and attestation service for seamless and verifiable deployment of middleboxes.  
We have also added several new features for commonly required functionalities: new specialized \click elements for secure packet processing, secure shared memory packet transfer for NFs chaining,  secure state persistence,  an efficient on-NIC timer for \sgx enclaves, and memory safety against \dpdk-specific Iago attacks. Furthermore, we have implemented several \sgx-specific optimizations in \slick. Our evaluation shows that \sys{} achieves near-native throughput and latency.



\end{abstract}

\author{Bohdan Trach$^\dag$, Alfred Krohmer$^\dag$, Sergei Arnautov$^\dag$, Franz Gregor$^\dag$,\\Pramod Bhatotia$^\ddag$, Christof Fetzer$^\dag$}
\affiliation{$^\dag$Technische Universit\"at Dresden \hspace{5mm} $^\ddag$University of Edinburgh}

\maketitle

\section{Introduction}
\label{sec:introduction}


Modern enterprises ubiquitously deploy network appliances or ``middleboxes" to manage their  networking infrastructure. These middleboxes  manage a wide-range of workflows for improving the efficiency (e.g., WAN optimizers), performance (e.g., caching, proxies),  reliability (e.g.,  load balancers  and monitoring), and security (e.g., firewalls and intrusion detection systems). Due to their wide-spread usage in the networking infrastructure, these   middleboxes incur significant deployment, maintenance, and management costs~\cite{aplomb}. 

To overcome these limitations, many enterprises are contemplating  to outsource the middlebox processing services in the cloud~\cite{aplomb, e2}. Cloud computing offers the economies of scale for computational resources with the ease of management, elasticity, and fault tolerance.  The realization of vision: ``middleboxes as a service in the cloud" is strengthened by the advancements in software-defined middleboxes, also known as Network Function Virtualization (NFV)~\cite{clickos}. Network Function Virtualization offers a flexible and modular architecture that can be easily deployed on the commodity hardware.  Thus, NFV is a perfect candidate to reap the outsourcing benefits of the (edge) cloud computing infrastructure.

However,  middleboxes that process confidential and private data cannot be securely deployed in the untrusted environment of the cloud. In the cloud environment, an accidental
or, in some cases, intentional action from a cloud administrator could compromise the confidentiality and integrity of execution. These threats of potential violations to the integrity and confidentiality of customer data is often cited as a key barrier to the adoption of cloud services~\cite{trusted-cloud-computing}.


To securely outsource middleboxes to the cloud, the state-of-the-art systems advocate network processing over the encrypted traffic~\cite{blindbox, embark}. However, these systems  support only restrictive type of middlebox functionalities, and incur prohibitively high performance overheads since they  require complex computations over the encrypted network traffic.


These limitations motivated our work---we strive to answer the following question: {\em How to securely outsource the middlebox processing service on the untrusted third-party platform without  sacrificing
performance while supporting a wide range of enterprise NFs?}

To answer this question, we propose \slick---a secure middlebox framework for deploying high-performance Network Functions (NFs) on untrusted commodity servers. The architecture of \slick is based on four design principles: (1) {\em Security} --- we aim to provide strong confidentiality and integrity guarantees for the execution of middleboxes against a powerful adversary, (2) {\em Performance} --- we strive to achieve near-native throughput and latency,  (3) {\em Generality} --- we aim to support a wide range of network functions (same as plain-text processing) with the ease of programmability, and, (4) {\em Transparency} --- we aim to provide a transparent, portable, and verifiable environment for deploying middleboxes.

\slick leverages \click~\cite{click} to provide a flexible and modular framework to build  a rich set of NFs using its out-of-the box elements and C++ extensions.  \slick leverages hardware-assisted secure enclaves based on \intel \sgx to provide strong confidentiality and integrity properties~\cite{sgx}. To achieve high performance despite the inherent limitations of the \sgx architecture, \slick builds on \scone~\cite{scone} (a shielded execution framework) and \intel \dpdk~\cite{dpdk} to efficiently process packets in the userspace secure enclave memory.  Finally, \slick builds on the container technology with a remote attestation interface~\cite{bootstrap-trust} to  provide a portable deployment and cryptographic verifiable mechanism.
 
Using the \sys framework, the middlebox owner launches a \sys instance in the cloud and performs remote attestation, passing \sys an encrypted configuration in case of successful attestation. Thereafter, \sys executes user-defined \click elements, which are responsible for reading packets in the userspace directly from NIC, performing network traffic processing, and sending them back to the network. All elements run inside \sgx enclave. Packets that must be processed under \sgx protection are copied into the enclave explicitly. We efficiently execute the expensive network I/O operations (to-and-from the enclave memory)  by integrating \scone with \dpdk.

 
Furthermore, we have designed several new design features for commonly required middlebox functionalities: (a) a remote attestation and configuration service for a seamless and verifiable deployment of middleboxes in the cloud, (b) new \click elements for secure packet processing, (c) an efficient and secure shared memory packet transfer in the multiple \sgx enclaves setup for NFVs chaining~\cite{nfv-chaining}, (d) a secure state persistence layer for fault-tolerance/migration~\cite{stateful-middleboxes},    (e) an on-NIC PTP clock as the time source for the \sgx enclaves, and (f)  a memory protection mechanism to defend against \dpdk-specific Iago attacks~\cite{iago-attacks}. 

We have implemented the aforementioned security features, and also added several \sgx-specific performance optimizations to \slick. Lastly, we have evaluated the system using a series of microbenchmarks, and two case-studies: a multiport IP Router, and Intrusion Detection System (IDS). Our evaluation shows that \sys{} achieves near-native throughput and latency. In order to improve throughput, we limit memory copying by storing most of the packets outside the enclave, using more efficient data structures and preallocating memory. In order to reduce latency, we use NIC timer when necessary, and avoid unnecessary system calls by modifying \click timer event scheduler.


%


\section{Background}
\label{sec:background}
In this section, we present a brief necessary background about the three technical building blocks of \slick: shielded execution, \dpdk, and \click.
\subsection{Shielded Execution}
\label{subsec:intel-sgx}
{\em Shielded execution} provides strong confidentiality and integrity guarantees for unmodified legacy applications running on untrusted platforms.
Our work builds on \scone~\cite{scone}---a shielded execution framework based on \intel \sgx~\cite{sgx}.

\myparagraph{Intel Software Guard Extensions (SGX)} \intel \sgx is a set of ISA extensions for Trusted Execution Environments (TEE) released as part of the Skylake architecture. \intel \sgx provides an abstraction of secure \emph{enclave}---a memory region for which the CPU guarantees the confidentiality and integrity of the data and code residing in it.
More specifically, the enclave memory is located in the Enclave Page Cache (EPC)---a dedicated memory region protected by  MEE, an on-chip Memory Encryption Engine. The MEE encrypts and decrypts cache lines with writes and reads in the EPC, respectively. The processor verifies that the  read/write accesses to the EPC are originated from the enclave code. Furthermore, the MEE verifies the integrity of the accessed page to detect memory modifications and rollback attacks.

However, the architecture of \sgx suffers from two major limitations: First, the EPC is a limited resource, currently restricted to 128~MB (out of which only ~94~MB is available to all enclaves). To overcome this limitation, \sgx supports a secure paging mechanism to an unprotected memory. However, the paging mechanism incurs very high overheads depending on the memory access pattern ($2\times$ to $2000\times$). Second, the execution of system calls is prohibited inside the enclave. To execute a system call, the executing thread has to exit the enclave. The system call arguments need to be copied in and out of the enclave memory. Such enclave transitions are expensive---especially, in the context of middleboxes---because of security checks and TLB flushes. 

\myparagraph{SCONE} SCONE is a shielded execution framework for unmodified applications based on Intel \sgx~\cite{scone}. In the SCONE framework, the legacy applications are statically compiled and linked against a modified standard C library (SCONE libc). In this model, application's address space is confined to the enclave memory, and interaction with the outside world (or the untrusted memory) is performed only via the system call interface. The SCONE libc executes system calls outside the enclave on behalf of the shielded application. The SCONE framework protects the executing application from the outside world, such as untrusted OS, through  \emph{shields}. In particular, shields copy arguments of system calls inside and outside the enclave and provide functionality to transparently encrypt the data that leaves the enclave perimeter. Furthermore, SCONE provides a {\em user-level threading} mechanism inside the enclave combined with the {\em asynchronous system call} mechanism in which threads outside the enclave asynchronously execute the system calls~\cite{flexSC}  without forcing the enclave threads to exit the enclave. 

\subsection{Intel DPDK and Click}



\myparagraph{Intel DPDK} We build on the \intel \dpdk library~\cite{dpdk} that supports developing high-performance networked systems running on commodity hardware. The \dpdk library allows processing of L2 packets from NIC directly in the userspace; thus it completely bypasses the OS networking stack to improve both the throughput and latency.

The \dpdk library consists of three main components: Environment Abstraction Layer (EAL), memory management subsystem, and Poll Mode Drivers (PMD). EAL provides a unified way to initialize the central \dpdk components (memory management,  poll drivers, threading, etc.). The memory management unit of \dpdk utilizes huge pages for the buffer management  through its own memory allocator  (\emph{mempool}). The \emph{mbuf} library provides functionality to store and process  the raw packet data (which is directly mapped in the userspace virtual memory) in the memory blocks allocated from a mempool.
Lastly,  a PMD uses the memory rings on the NIC to directly send and receive packets without interrupts, thus achieving high CPU utilization.



\myparagraph{Click} In addition to \dpdk, we leverage \click's~\cite{click} programmable and extensible architecture for the implementation of NFs using its out-of-the-box elements and C++ extensions. In particular, \click provides a dataflow programming language that allows construction of middleboxes as a graph of \emph{elements}---small, reusable, atomic pieces of network traffic processing functionality. Thereafter, \click routes packets through the elements in the dataflow graph. The modular architecture of \click greatly improves the programmability, productivity, and dependability of middleboxes.

\section{Overview}
\label{sec:overview}

\begin{figure}[t]
\centering

\includegraphics[scale=0.33]{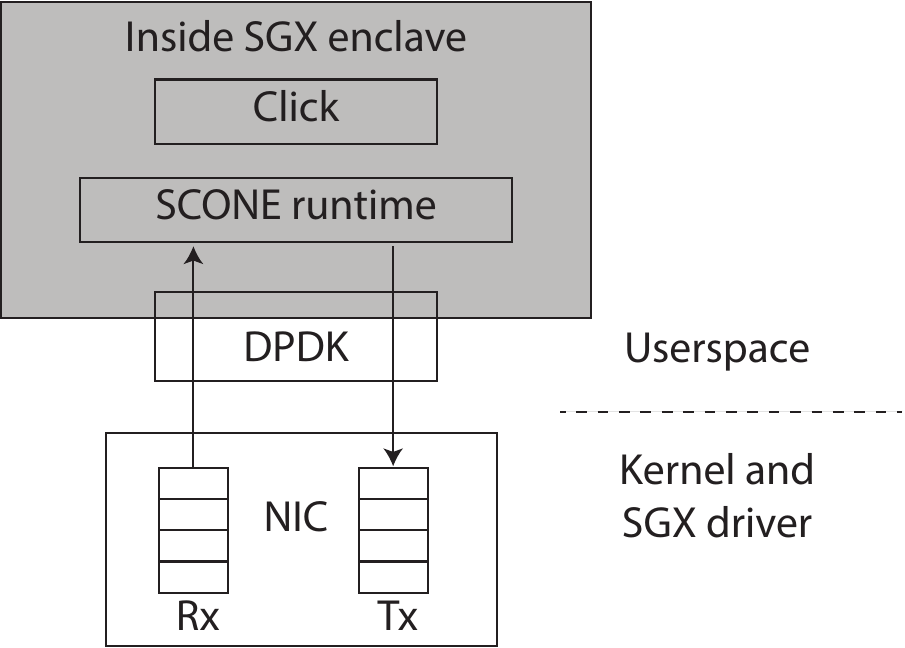}

\caption{\slick basic design}

\label{fig:basic-design}
\end{figure}

\myparagraph{Basic design} At a high-level, the core of our system \slick consists of a simple integration of a \dpdk-enabled \click that is running inside the \sgx enclave using \scone. Figure~\ref{fig:basic-design} shows the high-level architecture of \sys. 

While designing \slick, we need to take into account the architectural limitations of \intel \sgx.   As described in $\S$\ref{subsec:intel-sgx}, an enclave context switch (or exiting the enclave synchronously for issuing system calls) is quite expensive in the \sgx architecture.  The \scone framework overcomes this limitation using an asynchronous system call mechanism~\cite{flexSC}. While the asynchronous mechanism is good enough for commonly used services like HTTP servers or KV stores---it can not sustain the traffic rates of modern middleboxes. Therefore, we decided to use the userspace \dpdk I/O library as a better fit for the \sgx enclaves to achieve high performance.

Furthermore, we need to ensure that the memory footprint of \slick code and data is minimal, due to several reasons: As described in $\S$\ref{subsec:intel-sgx}, enclaves that use more than $94$MB of physical memory suffer high performance penalties due to EPC paging ($2\times$ to $2000\times$). In fact, to process data packets at line rate, even stricter resource limit must be obeyed---the working set of the application must fit into the L3 cache. Therefore, our design diligently ensures that we incur minimum cache misses,  and avoid EPC paging.

Besides performance reasons, minimizing the code size inside the enclave leads to a smaller Trusted Computing Base (TCB) and allows to reduce the attack surface. The core of \click is already quite small ($6$MB for a statically linked binary section that is loaded in the memory). We decrease its size by removing the unnecessary \click elements at the build time. Importantly, we designed \slick with the packet-related \dpdk data structures running outside of the enclave. 


More specifically, when the \scone runtime starts the application, it automatically places the application code, statically allocated data, and heap in the \sgx-protected memory. This mechanism is in contrast the way \dpdk operates---it 	by default allocates memory using {\tt x86\_64} huge page mechanism, which reduces the TLB miss rate. Such pages are not supported inside the enclave; besides that, NIC can only deliver packets to the unprotected memory, and network traffic entering or leaving machine can be modified by attacker. Therefore, we keep the huge pages enabled in \dpdk outside the enclave, and explicitly copy packets that must be processed with \sgx protection into the enclave. With this scheme, \dpdk-created packet data structures are allocated outside the \sgx enclave. We support an efficient data transfer between the \dpdk and enclave and processing inside the enclave using the new secure \click elements (detailed in $\S$\ref{subsec:secure-elements}).


When \sys starts, it performs remote attestation and obtains the configuration. \slick initializes the \dpdk subsystems, allocates huge page memory and takes the control over NICs that are available. Then, it starts running \click element scheduler, which reads packets from the NIC and passes them along the processing chain until they leave the system or are dropped. Packets that need \sgx protection are explicitly copied into enclave.



\myparagraph{Threat model}
\label{subsec:threat-model}
We target a scenario where the middleboxes that process confidential and  sensitive data are deployed in the untrusted cloud environment (or at the edge computing nodes at ISPs)~\cite{aplomb}.  Although the cloud providers are usually contractually obligated to not interfere with the compute and storage units, an accidental leakage, or even a malicious cloud operator might compromise the confidentiality and integrity of the execution~\cite{trusted-cloud-computing}. In the context of middleboxes~\cite{embark,blindbox}, attackers might try to learn the contents of the encrypted data packets and configuration of the system such as cryptographic keys, filtering and classification rules, etc. Furthermore, attackers might try to compromise the integrity of a middlebox by subverting its execution.

To circumvent such attacks, we protect against a very powerful adversary even in the presence of complex layers of software in the virtualized cloud computing infrastructure. More specifically, we assume that the adversary can control the entire system software stack, including the OS or the hypervisor, and is able to launch physical attacks, such as performing memory probes.

However, we note that \slick is not designed to protect against side-channel attacks~\cite{sgx-side-channels}, such as exploiting timing and page fault information. Furthermore, since the underlying infrastructure is controlled by the cloud operator we cannot defend against the denial-of-service attacks.

 



\myparagraph{System workflow}  Figure~\ref{fig:basic-workflow} shows the system workflow of \sys.
As a preparation to the deployment, developers build middlebox container images, and upload them to an image repository (such as Docker Hub~\cite{docker-hub}) using the \slick toolchain. Network operator, on the other hand, must bootstrap a Configuration and Attestation Service (CAS) on a trusted host, and a Local Attestation Service (LAS) on the host that will be running the middlebox. After this, \slick can be installed on the target machine in the cloud using the container technology---either manually or deployed as a container image from the image repository. Alternatively, it can be installed by transferring a single binary to the target machine.

The \slick framework is bootstrapped using the Configuration and Remote Attestation Service (CAS). The CAS service is launched either inside an \sgx enclave of a (already bootstrapped) untrusted machine in the cloud or on a trusted machine under the control of the middlebox operator outside the cloud. Middlebox developers implement the necessary NFs as \click configurations and send them to the CAS service together with all necessary secrets.

Once the operator launched \slick in the cloud, it will connect to CAS and carry out the remote attestation (detailed in $\S$\ref{subsec:cas}). If the remote attestation is successful, the \slick instance receives the configuration and necessary secrets. After that, \slick can start processing the network traffic, with all the benefits:

\begin{figure}[t]
\centering

\includegraphics[width=\columnwidth]{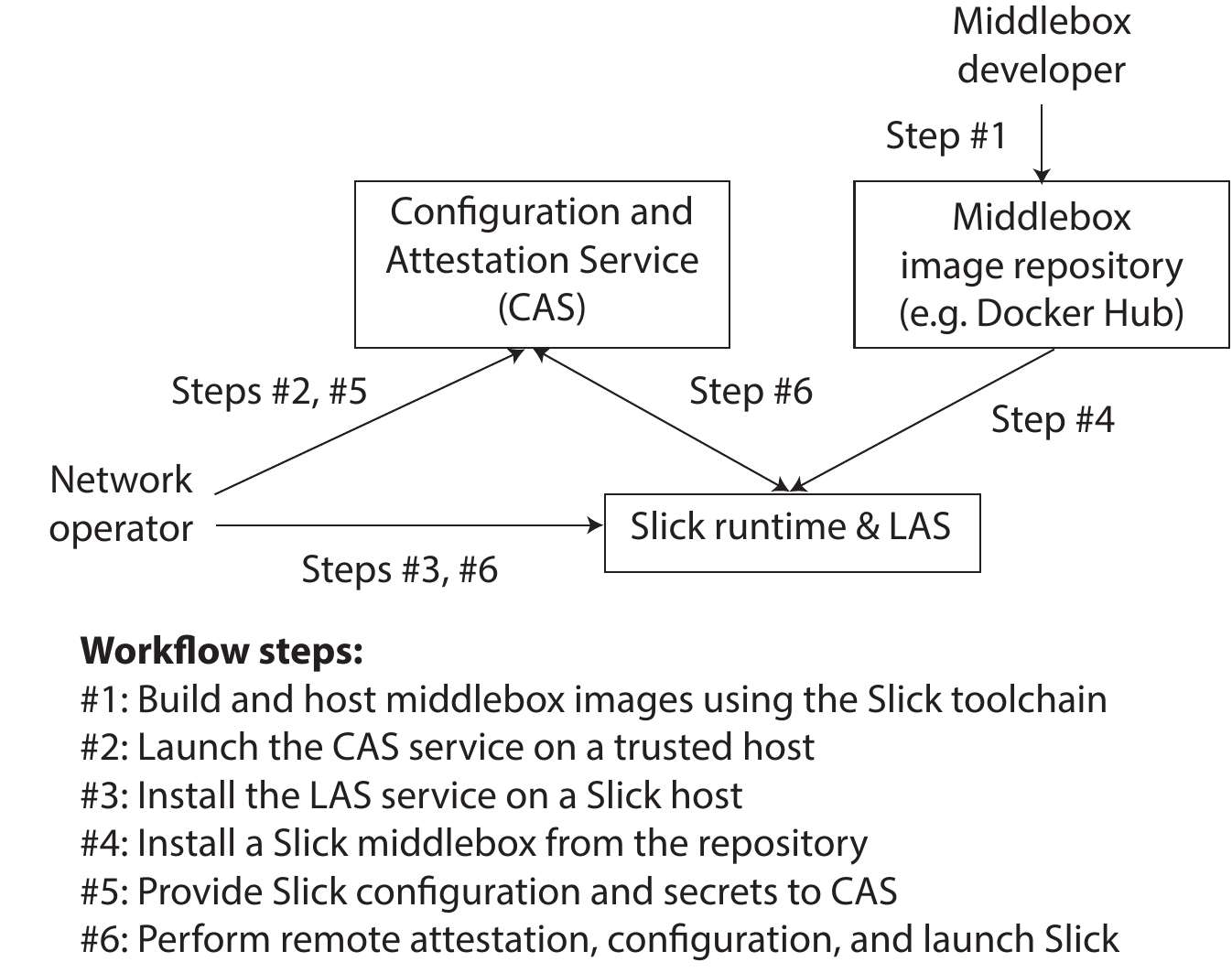}

\caption{\slick system workflow}

\label{fig:basic-workflow}
\end{figure}

\begin{figure*}[t]
\centering

\includegraphics[scale=0.7]{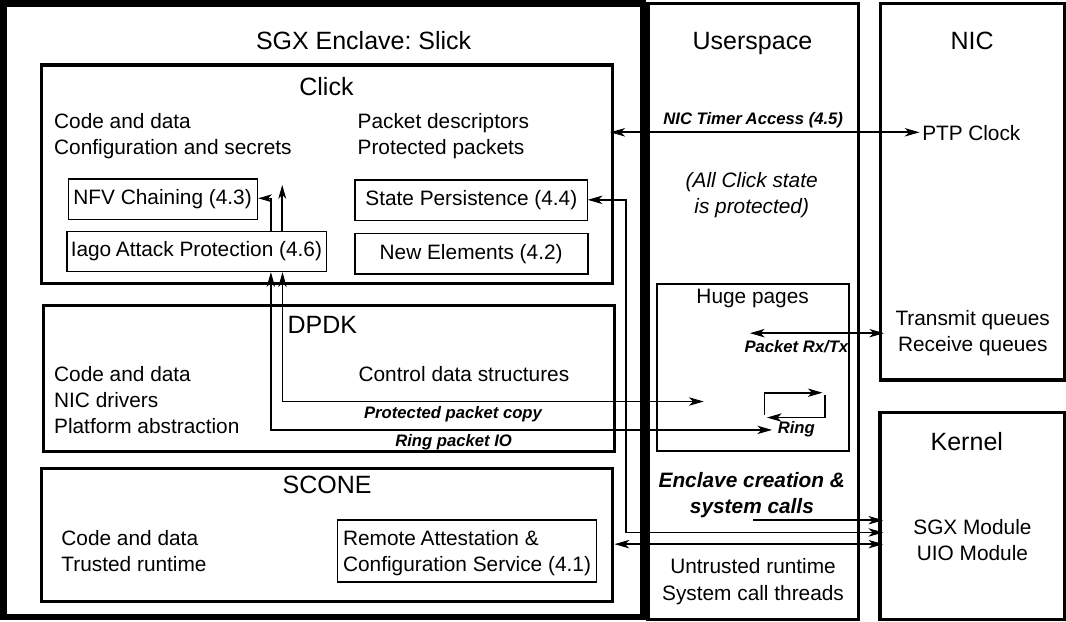}

\caption{\slick detailed design}

\label{fig:detailed-design}
\end{figure*}

\begin{itemize}

\item {\bf Security:} \slick provides strong confidentiality and integrity for the middlebox execution. We leverage hardware-assisted \sgx memory enclaves to provide strong security properties.

\item {\bf Performance:}  \slick achieves near-native throughput and latency by building a high-performance networking I/O architecture for the shielded execution of middleboxes. We overcome the architectural limitations of \intel \sgx enclaves by optimizing the combination of \scone and \dpdk.

\item {\bf Generality:} \slick supports a wide-range of NFVs, same as supported in the plain-text network processing, without restricting any functionalities. We leverage \click to provide a general framework to implement a wide-range of NFs.

\item {\bf Transparency:} Lastly, \slick provides network operators a portable, configurable, and verifiable architecture for the middlebox deployment in cloud. We build on the container technology with configuration and attestation services for a seamless deployment. 
\end{itemize}

\myparagraph{Limitation} We note that neither \dpdk nor \click support flow-based stateful traffic. This implies that \slick currently supports NFs that work on L2 and L3; as only restricted processing of L4-L6 traffic is supported. We plan to extend our system to support flow-based traffic with shielded execution by integrating a user-level networking stack~\cite{mTCP}.

\section{Design Details}
\label{sec:advanced}
We next present the design details of \slick. Figure~\ref{fig:detailed-design} shows the detailed architecture of \slick.

\subsection{Configuration and Remote Attestation}
\label{subsec:cas}

\begin{figure}[t]
\centering

\includegraphics[width=\columnwidth]{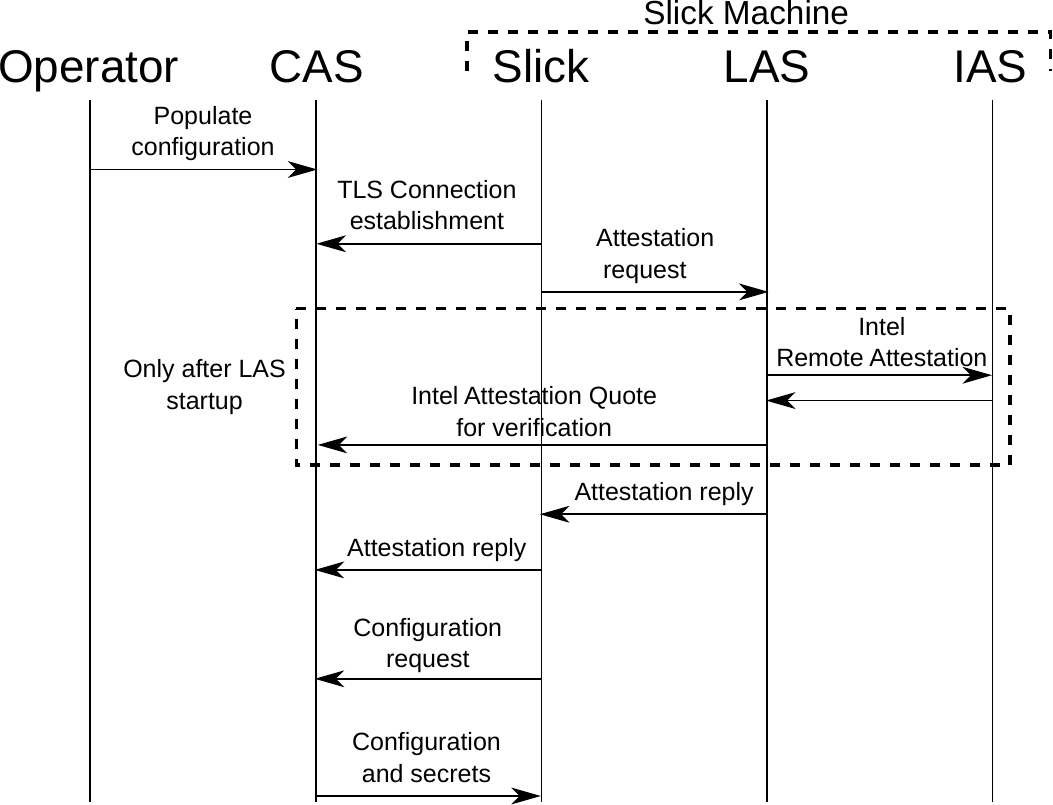}

\caption{\slick's configuration and attestation}

\label{fig:cas-workflow}
\end{figure}

To bootstrap a trusted middlebox system in the cloud, one has to establish trust in the system components. While \intel \sgx provides a remote attestation feature, a holistic system must be built for remote attestation and secure configuration of middleboxes~\cite{excalibur}. To achieve this goal, we have designed a Configuration and Attestation Service (CAS) for middleboxes. Figure~\ref{fig:cas-workflow} shows the system protocol for the CAS service.


In order to attest an enclave using Intel Remote Attestation, verifier (operator of a \slick instance) connects to the application and requests a quote. The enclave requests a report from SGX hardware and transmits it to the Intel Quoting Enclave (QE), which verifies, signs, and sends back the report. The enclave then forwards it to the verifier. This quote can be verified using the Intel verification service~\cite{ias}.

Our remote attestation system extends Intel's RA service, and is integrated with a configuration system, which provisions \slick with its configuration in a secure way using a trusted channel established during  attestation. This system consists of enclave-level library, Local Attestation Service (LAS), and Configuration and Attestation Service (CAS).
\begin{itemize}
\item Enclave library interacts with LAS and CAS to carry out remote attestation, and allows setting environment variables, command-line arguments, and keys for the \scone shielding layer in a secure and confidential manner.
\item Local Attestation Service is running on the same machine as \slick middlebox, and acts as a proxy for interaction with the Intel Attestation Service (IAS). It also acts as the intermediate root of trust: once LAS is attested, further \slick instances can be launched even when IAS is unavailable.
\item Configuration Attestation Service is running on a single (possibly replicated) node and stores configuration and secrets of the services built with \scone. It maintains information about attested enclave instances, and provisions configuration to applications using the Enclave Library.
\end{itemize}

To bootstrap the system, the operator launches CAS, either on the host under his control or on the host in the cloud inside an SGX enclave. Then, the CAS service is populated with configurations and secrets using the REST API or a command-line configuration tool. LAS and IAS instances are launched on cloud hosts that will run \slick instances. During startup, each \slick instance establishes a TLS connection to CAS. Simultaneously, it connects to LAS to perform local attestation. In the case the LAS instance is not yet attested, it will attest itself to CAS using Intel Remote Attestation protocol, and then attest \slick instance to CAS using SGX local attestation. Local attestation verifies integrity of the \slick binary, and establishes whether \slick is running under SGX protection. This allows to remove distribution mechanisms (such as Docker Hub) from the TCB. After that, \slick sends the LAS quote to CAS, and a secure channel is established between CAS and \slick. Thereafter, \slick obtains its configuration from the CAS service and transfers control to main \slick code.


\subsection{Secure Elements}
\label{subsec:secure-elements}

 As described in $\S$\ref{sec:overview},  we designed \slick with the packet-related data structures of \dpdk running outside the enclave. Therefore, we needed an efficient way to support the communication between \dpdk and the enclave memory region. In particular, we have to consider the overheads of accessing the SGX-encrypted pages from the main memory and copying of the data between the protected and unprotected memory regions. When possible, the data packets with plain-text contents should not be needlessly copied into the enclave, as it will degrade the performance. Therefore, we designed specialized secure \click elements (shown in Table~\ref{tab:new-elements})  for copying the data packets into/outside the enclave to facilitate efficient communication. 

By default, packets are read from NIC queues into the untrusted memory. This reduces the overhead of using \sgx when processing packets that are not encrypted and can be safely treated with less security mechanisms involved. Such packets are immediately forwarded or dropped upon header inspection. On the other hand, we must move packets into the enclave memory with explicit copy element. We have implemented such an element ({\tt ToEnclave}), and use it to construct secure packet processing chains.

We have also added support for the commonly used AES-GCM cipher into \slick ({\tt Seal} and {\tt Unseal} elements). This allows us to construct VPN systems that use modern cryptographic mechanisms. This element was implemented using Intel ISA-L crypto library. In order to allow secure key generation inside the enclave, we have exposed \scone functions for getting \sgx {\tt Seal} keys to the  \slick internal APIs.

To allow building high-performance IDS systems based using \slick, we have created and element based on the {\tt HyperScan} regular expression library. It allows fast matching of a number of regular expressions for the incoming packets, simplifying implementation of systems like Snort~\cite{snort}.

We have also added elements that implement more broad mechanisms: {\tt DPDKRing} ($\S$\ref{subsec:nfv-chaining}) for NFV chaining, and {\tt StateFile} ($\S$\ref{subsec:stateful}) for state persistence in middleboxes.

\subsection{NFVs Chaining}
\label{subsec:nfv-chaining}

Typically NFVs are chained together to build a dataflow processing graph with multiple \click elements, spanning across multiple cores, sockets, and machines~\cite{e2, nfv-chaining}. The communication between different cores and sockets happens through the shared memory, and communication across machines via NICs over RDMA/Infiniband. \dpdk supports NUMA systems and always explicitly tries to allocate memory from the local socket RAM nodes.

However, unlike normal POSIX applications, \sgx enclaves can not be shared across different sockets. (The future \sgx-enabled servers might have support for the NUMA architecture.) As a result, in the current \intel \sgx architecture, the users would need to run one \slick instance per each CPU socket. Another reason for cross-instance chaining is collocation of middleboxes from different developers that do not necessarily trust each other. In this case, developers would want to leverage SGX to protect the secrets. Therefore,  it is imperative for the \slick framework to provide an efficient communication mechanisms between enclaves to support high-performance NFVs chaining.

We built an efficient mechanism for communication between different
\slick instances by leveraging existing \dpdk features. In particular,
\dpdk already provides a building block for high performance
communication between different threads or processes with its ring API.
This API contains highly optimized implementations of concurrent,
lockless FIFO queues which use huge page memory for storage. We have
implemented the {\tt DPDKRing} element (see Table~\ref{tab:new-elements}) for \slick to utilize it for chaining. As huge
page memory is shared between multiple \slick instances, the ring buffers
are shared as well and can be used as an efficient way of communication
between multiple \slick processes.

\begin{table}[t]\small

\setlength\tabcolsep{0pt}
\centering
\begin{tabular}{p{8.0cm}}

\hline\\[-8pt] 
{\tt ToEnclave} \\
\hspace*{0.5cm} Transfers a packet to enclave, frees the original packet\\
\\[-8pt] 
{\tt Seal(Key, Security Association state)} \\
\hspace*{0.5cm} Encrypts the packet with AES-GCM\\
\\[-8pt] 
{\tt Unseal(Key, Security Association state)} \\
\hspace*{0.5cm} Decrypts the packet with AES-GCM\\
\\[-8pt] 
{\tt HyperScan(rule database)} \\
\hspace*{0.5cm}  High-performance regular expression matching engine \\
\\[-8pt]
{\tt DPDKRing(Ring name)} \\
\hspace*{0.5cm} Transfers a packet to the DPDK ring structures\\
\\[-8pt]
{\tt StateFile(Key, path)} \\
\hspace*{0.5cm} Provides settings to the persistent state engine\\
\hline
\end{tabular}

\caption{\slick new specialized elements}\label{tab:new-elements}
\end{table}

This solution requires assigning ownership of all shared data structures
to a single process. For this, we rely on \dpdk distinction between primary
and secondary processes. Primary processes,
the default type, request huge page memory from the operating system,
allocate memory pools and initialize the hardware. Secondary processes
skip device initialization and map the huge page memory already
requested by the primary process into their own address space. 
In order to support NFV chaining using multiple processes on the same machine, we
added support for starting \slick instances as secondary \dpdk processes.

Depending on the process type, the {\tt DPDKRing} element either creates a new
ring (primary process) or looks up an existing ring (secondary process)
in huge page memory. In \slick, packets pushed towards a DPDKRing element
are enqueued into the ring and can be dequeued from the ring in another
process for further processing. A bidirectional communication between
two processes can be established by using a pair of rings. 
\begin{table}[t]\small

\setlength\tabcolsep{0pt}
\centering
\begin{tabular}{p{8.0cm}}

\hline\\[-8pt]
{\tt Seal(StateFile)} \\
\hspace*{0.5cm}Seals elements' state in the {\tt StateFile} \\
\\[-8pt] 
{\tt Unseal(StateFile)} \\
\hspace*{0.5cm}Unseals elements' state from the {\tt StateFile}\\
\\[-8pt]
{\tt Persist(timer, StateFile)} \\
\hspace*{0.5cm} Periodically persists the state to {\tt StateFile} \\
\hline
\end{tabular}
\caption{\slick APIs for state persistence}\label{tab:stateful}
\end{table}
\subsection{Middlebox State Persistence}
\label{subsec:stateful}

Middleboxes often maintain useful state (such as counter values, Ethernet switch mapping, activity logs, routing table, etc.) for fault-tolerance~\cite{stateful-middleboxes}, migration~\cite{middlebox-migration}, diagnostics~\cite{middlebox-perf}, etc. To securely persist this state, we extend \slick with new APIs (shown in Table~\ref{tab:stateful}) for the state persistence in middleboxes. The {\tt Seal} primitive is used to collect the state that must be persisted from the elements, and write it down in encrypted form to disk. {\tt Unseal} reads this state from disk, decrypts it and populates the elements with this state.



To configure this functionality, we have added a new configuration element to \slick, called {\tt StateFile} (see Table~\ref{tab:new-elements}). Its parameters are file to which state should be written and the key that should be used for encryption. Note that this information is transmitted to \slick instance in the configuration string via remote attestation, and is not accessible outside the enclave. We don't use \scone file system shield, and instead encrypt and decrypt file as a single block. This ensures confidentiality and integrity of stored data via the use of AES-GCM cipher. 

We do not attempt to extract the relevant state transparently. Instead, we rely on programmer providing necessary serialization routines that save only necessary parts of element state. These routines are available in \slick as read and write handlers, and are triggered in the \slick startup procedure after the configuration is loaded, parsed, and initialization of the basic components  is finished, or manually via {\tt ControlSocket} interface of the {\tt StateFile} element. It's also possible to trigger them periodically via a timer. 

%
%


\subsection{NIC Time Source}
\label{subsec:nic-timer}
Timer is one of the commonly used functionalities in middleboxes~\cite{clickos, e2}. It is used for a variety of purposes such as measuring performance, scheduling execution of periodically-triggered NFs, etc. 

The time measurement can be fine-grained or coarse-grained based on the application requirements. For the fine-grained cycle-level measurements, developers use {\tt rdtscp} instruction, which is extremely cheap and precise. Whereas for the coarse-grained measurements, applications invoke system calls like {\tt gettimeofday} or {\tt clock\_gettime}.

However, in the context \sgx enclaves, both {\tt rdtsc} and system calls have unacceptable latency for their use in middleboxes to process the network traffic at line rate. More specifically,  the {\tt rdtscp} instruction is forbidden inside the enclave, and therefore, it causes an enclave exit event; whereas, asynchronous system calls in \scone are submitted though a system call queue that is optimized for the raw throughput, but not latency.

To overcome these limitations, we have opted to use the on-NIC PTP clock as the clock source for the enclave. This clock can be read inside the enclave reasonably fast ($0.9$ $\mu$sec, which is on the same scale of magnitude as reading HPET). Moreover, it neither causes enclave exits nor requires submitting system calls. Furthermore, the on-NIC clock is extremely precise since it is intended to use for the PTP synchronization protocol. 

We stress that this time source is not secure, and can be used as a denial of service attack vector by the malicious operation system. However, the same is true for the other time sources---a trusted, efficient and precise time source for \sgx enclaves remains an unsolved problem that will likely require changes to the CPU hardware~\cite{malware-guard-extensions}.

\subsection{DPDK-Specific Iago Attacks}
\label{subsec:iago-attack}

Iago attacks~\cite{iago-attacks} are a serious class of security attacks
that can be launched on shielded execution to compromise the
confidentiality and integrity properties. In
particular, an Iago attack originates through malicious inputs supplied
by the untrusted code to the trusted code. In the classical setting, a malicious
OS can subvert the execution of an \sgx-protected application by exploiting
the application's dependency on correct values of system
call return values~\cite{haven}.

The decision ($\S$\ref{sec:overview}) to allocate huge pages for
packet buffers and \dpdk rings has security implications. The fact that
packets are passed through rings by reference, and \dpdk buffers contain
pointers, opens a new attack surface on \slick. Attackers with access to
this memory region could modify pointers to point into the SGX-protected
regions and make the enclave inadvertently leak secrets over the network. 

To protect against \dpdk-specific Iago attacks, we have
implemented a pointer validation function. More specifically, the scheme
uses an enclave parameter structure that is located inside the enclave
memory and defines the enclave memory boundaries. Pointers are validated by
checking if they do \emph{not} point into the enclave memory range
$[base, base+enclave\_size)$. We note that \slick is already protected against the
classical syscall-specific Iago attacks through \scone's {\em shielding}
interfaces.

This ensures that no pointers possibly pointing to the secrets stored in EPC
are accepted through the unprotected huge page memory. Pointers can
still be modified by a malicious attacker, but they can only point to
the unprotected memory. 

As it is possible for an application to enqueue and dequeue arbitrary
pointers into \dpdk's {\tt rte\_ring} structures,
it is not easily possible to integrate this pointer check directly into
\dpdk. Instead, we implemented these pointer checks in the {\tt DPDKRing} and
{\tt FromDPDKDevice}  ($\S$\ref{subsec:nfv-chaining}) elements. If \slick detects a malicious
pointer, it assumes an attack, notifies the application operator and
drops the packet.

\section{Implementation}
\label{sec:implementation}

\subsection{Toolchain}
We built \slick's toolchain using \dpdk (version 16.11) and \click (master branch commit {\tt 0e860a93}). We further integrate it with the \scone runtime to compile \slick. The toolchain is based on the {\tt musl-cross-make}~\cite{musl-cross-make} project, modified to use \scone{} {\tt libc} instead of the standard {\tt musl-libc}. We use {\tt gcc} version $6.3.0$ for the compilation process. This enables us to use both {\tt C} and {\tt C++} code. {\tt musl-cross-make} is also used to compile native version of \click for the evaluation.  We used Boost {\tt C++} library (version $1.63$) to build a static version of the Hyperscan high performance regular expression matching engine (master branch commit {\tt 7aff6f61}) and incorporated it into \slick. We use {\tt WolfSSL} library~\cite{wolfssl} to implement the remote attestation system and {\tt StateFile} sealing, and packet {\tt Seal/Unseal} elements. The toolchain contains automated scripts for building and deploying middlebox container images, and setting up \slick and CAS services (as described in the system workflow in $\S$\ref{sec:overview}).

To make the compilation of \slick work with the \scone toolchain, some changes to \dpdk were necessary. In particular, we need to remove the helper functions for printing stack tracebacks and provide some glibc-specific structures and macros as well as kernel header files. \click is implemented in {\tt C++} using mostly high level APIs and required no adaptions. 

The resulting \slick binary is $8.2$ MB in size, and around $16$ MB including minimal runtime stack and heap allocation. This implies that we could run roughly up to  six instances of \slick in parallel on one processor without impacting the performance by EPC paging (system-wide EPC limit is 94 MB).

\subsection{Memory Management} \scone implements it's own, in-enclave memory management, using EPC pages for all {\tt malloc} and {\tt mmap} anonymous memory allocation calls. Allocating huge page memory through this allocator is not possible without modifications.

However, accessing huge pages in \dpdk does not necessarily require bypassing \scone, because of the specific way \dpdk uses to allocate huge pages. Instead of passing {\tt MAP\_HUGETLB} flag to {\tt mmap}(), it opens shared memory files in the {\tt hugetlbfs} virtual filesystem and passes those file descriptors to {\tt mmap} call, which is not protected by \scone. This {\tt mmap} file-to-memory mapping request is directly passed to the operating system by \scone.

\subsection{Optimizations}
In order to further improve the performance, especially for the case of \dpdk running inside the enclave, we optimized the data path inside \click. We used the Linux performance profiling tool {\tt perf}~\cite{perf} to find the bottlenecks.

\myparagraph{Memory pre-allocation} The {\tt FromDPDK} element allocated memory for packet descriptor storage on the stack each time the {\tt run\_task} function was called. We pre-allocated this memory once in a constructor.

\myparagraph{Branching hints} We inserted GCC-specific {\tt unlikely} / {\tt likely} macros in several {\tt if}-clauses. These get translated to platform specific instructions which instruct the processor to always try the given branch first instead of using its built-in branch prediction.

\myparagraph{Modulo operations} We replaced all modulo operations in the data path by compare-and-subtract operations (which are way less expensive in terms of processing time).

\myparagraph{Queue optimization} In the {\tt ToDPDKDevice} \click element we replaced the inefficient implementation of the queue (which used {\tt std::vector} from the {\tt C++} standard library) by the {\tt rte\_ring} structure provided by \dpdk.

\myparagraph{Timer event scheduler optimization} In the \click timer event scheduler, we have optimized the code to reduce the amount of {\tt clock\_gettime} system calls. This allowed us to reduce  the  latency in short element chains to the native level.


\section{Evaluation}
\label{sec:evaluation}
\subsection{Experimental Setup}

\myparagraph{Testbed} We benchmark our system using two machines: (1) load generator, and (2) \sgx-enabled machine. The load generator is a Broadwell Intel Xeon D-1540 (2.00GHz, 8 cores, 16 hyperthreads) machine with $32$ GB RAM. The \sgx machine under test is Intel Xeon E3-1270 v5 (3.60GHz, 4 cores, 8 hyper-threads) with $32$ GB RAM running Linux kernel 4.4. Each core has private $32$KB L1 and $256$KB L2 caches, and all cores share a $8$MB L3 cache. The load generator is connected to the test machine using $40$ GbE Intel XL-710 network card. We use {\tt pktgen-dpdk} for throughput testing. The test machine is configured to use the maximum number of cores. The load generator saturates the link starting with $128$ byte packets.

\myparagraph{Applications} For the microbenchmarks, we use three basic \click elements: (1) {\tt Wire}, which sends the packet immediately after receiving; (2) {\tt EtherMirror}, which sends the packet after swapping the source and destination addresses; and (3) {\tt Firewall}, which does packet filtering based on PCAP-like rules. The {\tt Firewall} element is configured with $10$ rules.

For the case-studies, we evaluated \slick using two applications: (1) a multiport {\tt IPRouter}, and (2) an {\tt IDS}.

\myparagraph{Methodology} For the performance measurements, we consider several cases of our system:
\begin{itemize}
\item Native (\sgx-independent) with and without generic optimizations.
\item \sgx-enabled \sys{} with and without optimizations.
\item \sgx-enabled \sys{} with the on-NIC timer.
\end{itemize}


\begin{figure}[t]
 \centering
 \includegraphics[width=\columnwidth]{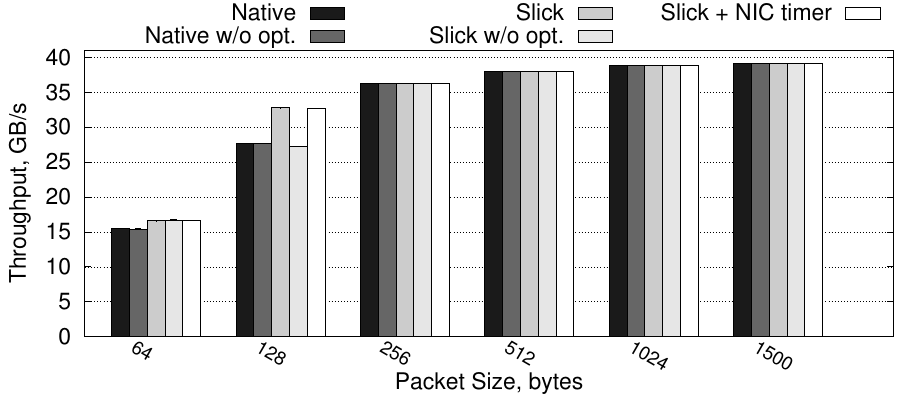}
 \caption{Wire throughput w/ varying packet size}
 \label{fig:throughput-packets-wire}
\end{figure}
\begin{figure}[t]
 \centering
 \includegraphics[width=\columnwidth]{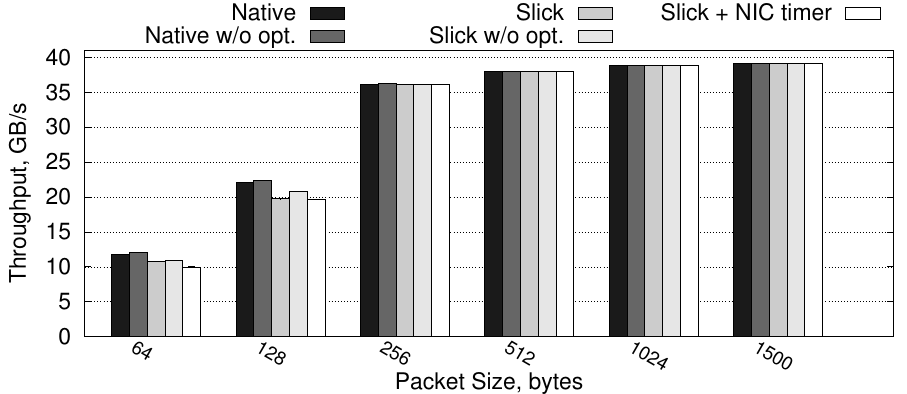}
 \caption{EtherMirror throughput w/ varying packet size}
 \label{fig:throughput-packets-ethermirror}
\end{figure}

\subsection{Throughput}
\label{sec:eval-throughput-basic}

\begin{figure}[t]
 \centering
 \includegraphics[width=\columnwidth]{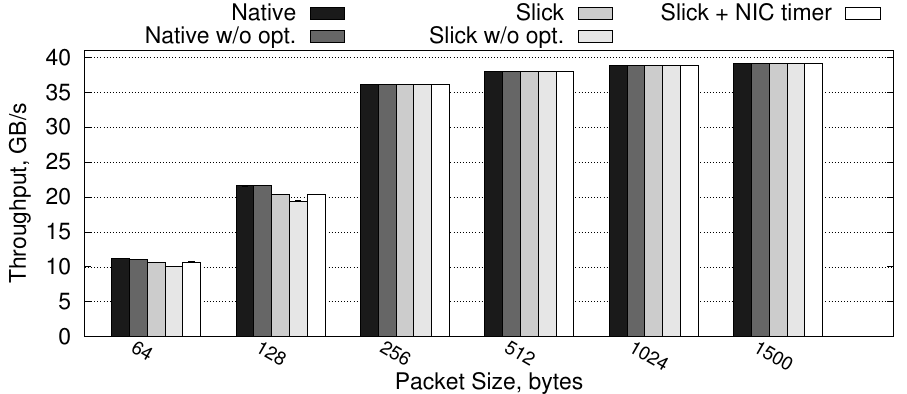}
 \caption{Firewall throughput w/ varying packet size}
 \label{fig:throughput-packets-firewall}
\end{figure}

We first measure the throughput of three microbenchmark applications with varying packet size. \sys is running on four cores. Figure~\ref{fig:throughput-packets-wire}, Figure~\ref{fig:throughput-packets-ethermirror}, and Figure~\ref{fig:throughput-packets-firewall} present the throughput for {\tt Wire}, {\tt Ethermirror}, and {\tt Firewall}, respectively.

The results show that the performance of \sys{} matches the performance of \click. In the case of Wire application with the packet sizes smaller than 256 bytes, \slick is better than the native version.  This is explained by the fact that \click timer event scheduler optimization is missing in the native \click, which removes some system call overhead from the {\tt Wire} application. The impact is smaller with other applications, because they contain elements that reduce the relative overhead of \click scheduler. We also see that \sys{} achieves line rate at $512$ byte packets. 

\subsection{Scalability}
\label{subsec:eval:scalability}

\begin{figure}[t]
 \centering
 \includegraphics[width=\columnwidth]{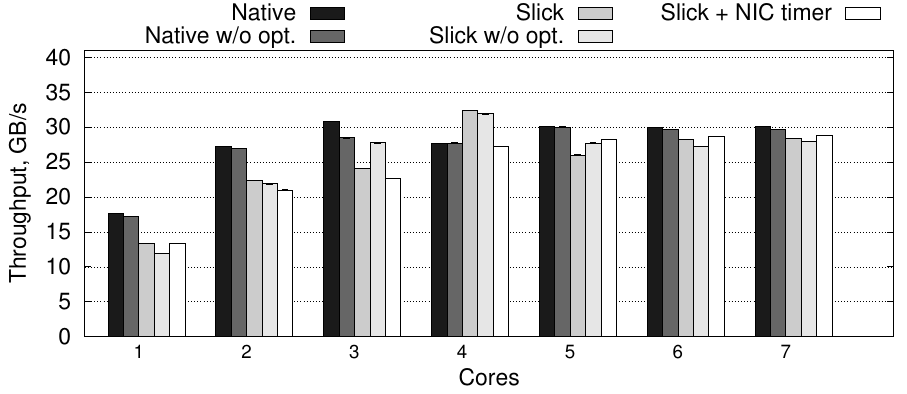}
 \caption{Wire throughput w/ increasing cores}
 \label{fig:throughput-packets-wire-core}
\end{figure}

\begin{figure}[t]
 \centering
 \includegraphics[width=\columnwidth]{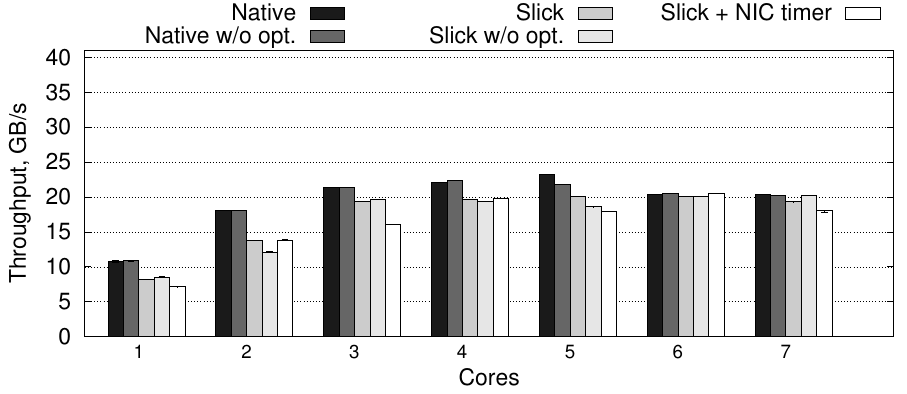}
 \caption{EtherMirror throughput w/ increasing cores}
 \label{fig:throughput-packets-ethermirror-core}
\end{figure}

\begin{figure}[t]
 \centering
 \includegraphics[width=\columnwidth]{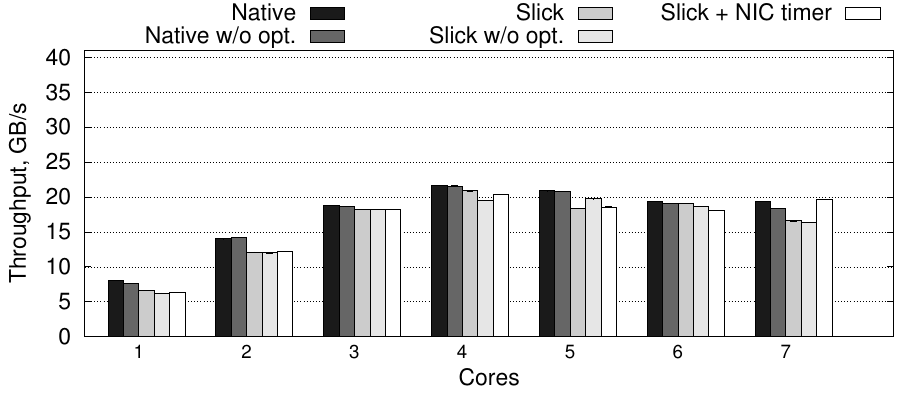}
 \caption{Firewall throughput w/ increasing cores}
 \label{fig:throughput-packets-firewall-core}
\end{figure}

We next consider the scalability of our system with increasing number of cores. (Our latest \sgx-enabled server has maximum of $4$ cores / $8$ HT. Recently released Intel X-Series will consist of $18$ cores.) Figure~\ref{fig:throughput-packets-wire-core}, Figure~\ref{fig:throughput-packets-ethermirror-core}, and Figure~\ref{fig:throughput-packets-firewall-core} present the throughput for {\tt Wire}, {\tt Ethermirror}, and {\tt Firewall}, respectively. The scalability of both \sys and \click is limited. We can see that the performance for both native and \sys peaks at four cores.  There are several reasons for this:
\begin{itemize}
\item \dpdk and \slick work best with hyperthreading disabled.
\item \sys runs system call threads to execute system calls asynchronously. When there are no system calls for a long time, they back-off (i.e. yield the CPU). They wake up from back-off periodically, and as result, they preempt in-enclave threads.
\end{itemize}

\subsection{Latency}
\label{sec:eval-latency}
\begin{figure}[t]
 \centering
 \includegraphics[width=\columnwidth]{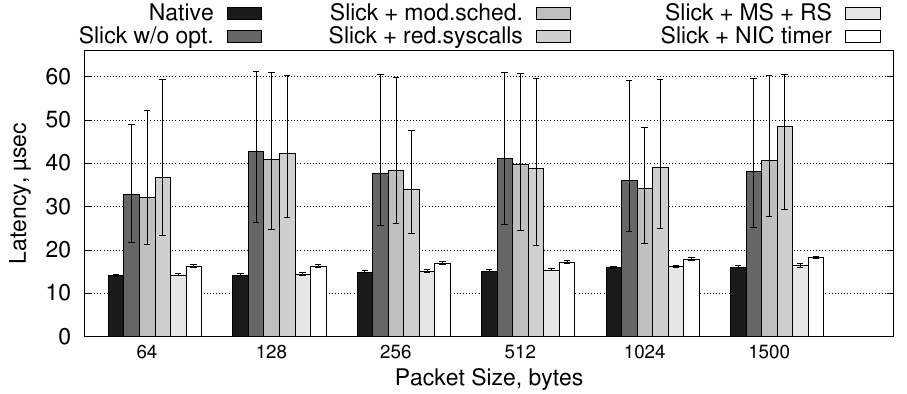}
 \caption{EtherMirror latency measurement}
 \label{fig:latency-ethermirror}
\end{figure}

To evaluate the influence of \slick's runtime environment on latency, we have also measured the packet processing latency using the following scheme: load generator runs an application, continuously generates a UDP packet and waits for its return spinning. We study packet round trip time measured at the load generator. On the \slick instance, we are running the {\tt EtherMirror} application. (We omit the results for other applications due to the space constraints.) 
For this measurements, we did not perform any latency-specific tuning of the environment other than thread pinning, which is enabled by default in \dpdk. We expect that a production system with stringent requirements to low latency will use \lstinline{SCHED_FIFO} scheduler and have isolated cores. Figure~\ref{fig:latency-ethermirror} present the latency measurements for {\tt EtherMirror}.

The low performance of \slick without optimizations is explained by the fact that \slick executes {\tt clock\_gettime} system calls in the timer event scheduling code. \scone system calls are optimized for  raw throughput with a large number of threads, but not for low latency; this makes the latency measurement result $3\times$ worse than the native execution. We have considered the following latency optimizations:
\begin{itemize}
\item Reduced system call rate for immediate\-ly-sche\-duled timer events. It removes one system call round-trip from the packet latency.
\item Modified scheduler that prioritizes immediately-sche\-duled events and allows to remove a system call from scheduler if there are no periodic timer events.
\end{itemize}


One of the surprising results that we have is that each of these optimizations does not have a statistically significant influence when applied individually, which can be explained by the fact that once the system call thread has left the back-off mode, it will execute system calls with low individual overhead. On the other hand, when applied simultaneously, they return the latency to almost-native levels---influence of \sgx and \scone on the latency is extremely small.

We consider using NIC timer as a separate optimization. One can see that reading NIC timer is a costly operation; it happens twice per packet in our measurements, adding approximately $0.9*2 = 1.8 \mu$sec to the total latency. On the other hand, it's much faster than executing clock-reading system calls, and can further improve system timeliness when combined with other optimizations.

\subsection{Configuration and Attestation}
\label{sec:eval-cas}

\begin{table}[t]\small
\setlength\tabcolsep{0pt}
\centering
\begin{tabular}{lc}
  \hline Phase & Average Duration, µsec \\
  \hline Attestation  & 19467 \\
  \hspace*{0.5cm}CAS communication & 19301 \\
  \hspace*{0.5cm}LAS communication & 1474 \\
  Configuration  & 825.6 \\
  Total time & 26368 \\
  \hline
\end{tabular}
\caption{Overheads of configuration and attestation}\label{tab:remote}
\end{table}


We next evaluate the overheads of the configuration and attestation service in \slick. The measurement results are presented in Table~\ref{tab:remote}. The results show that remote attestation has negligible effect on the \slick startup time. Furthermore, even though TLS session establishment is a costly operation, it is performed once per instance start-up, allowing operator to use a single CAS node for thousands of \slick instances.


\subsection{ToEnclave Overhead}
\label{sec:eval-secure-elements}
\begin{figure}[t]
 \centering
 \includegraphics[width=\columnwidth]{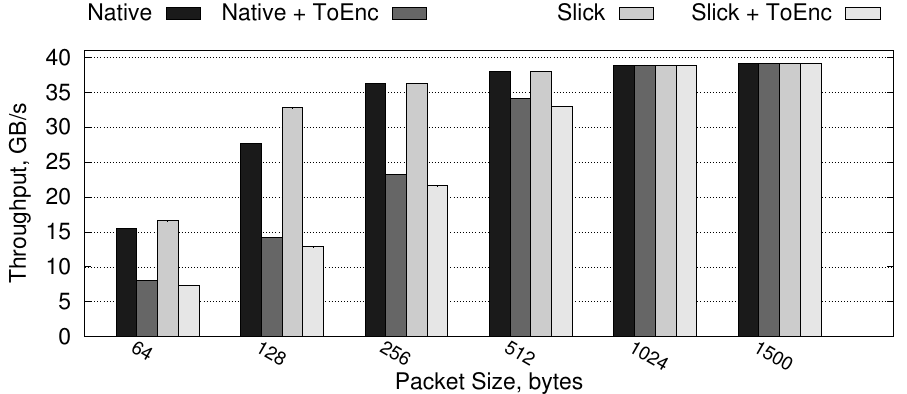}
 \caption{Throughput w/ ToEnclave + Wire}
 \label{fig:toenclave-wire}
\end{figure}

\begin{figure}[t]
 \centering
 \includegraphics[width=\columnwidth]{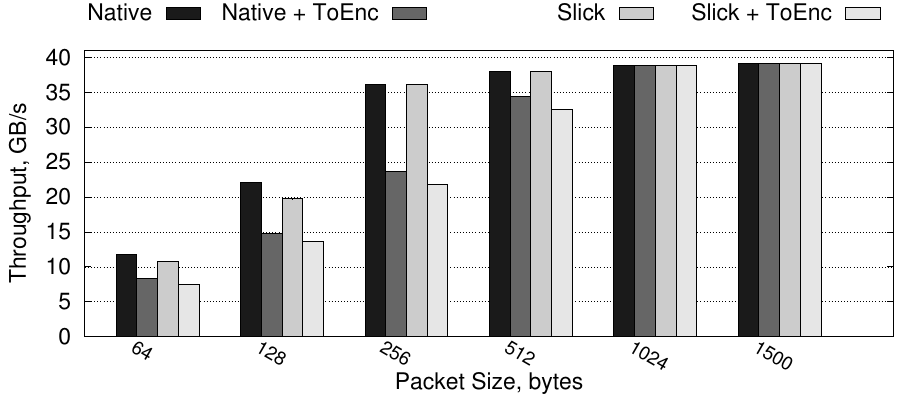}
 \caption{Throughput w/ ToEnclave + EtherMirror}
 \label{fig:toenclave-ethermirror}
\end{figure}

We next measure the throughput of the new secure {\tt ToEnclave} element added in \slick, which is used to copy the packet data inside SGX enclave protected memory. We evaluate the impact of this extra data copy by measuring the throughput scaling with varying packet size. Figure~\ref{fig:toenclave-wire} and Figure~\ref{fig:toenclave-ethermirror} show the results for {\tt Wire} and {\tt EtherMirror}, respectively.

As we can see that the overhead of the extra memory copy peaks with small packet sizes. This is due to the fact that for each received packet, operations with rather high overhead must be executed to allocate the packet. One way to reduce this cost would be to batch the memory allocation for all packets. It can be also seen that the overhead with \slick when compared to the native execution is relatively small: \slick with {\tt ToEnclave} is running within 88\% of the native version with extra memory copy in the worst case of small packet sizes, and within 60\% of the native Click without {\tt ToEnclave} element.

\subsection{NFVs Chaining}
\label{subsec:eval:chaining}
To measure the throughput of the NFV chaining scheme, we have implemented a chaining application. The chaining application implements packet communication between two \slick instances running on the same machine through a \dpdk packet ring. One instance contains an application that receives packets from network and sends them to the other instance via the {\tt DPDKRing} element. The second instance receives packets from the ring and sends them back through a different {\tt DPDKRing} element. These packets are received by the first \slick forming a circular ring. Thereafter, the packets are transmitted back to the load generating node. Please note that the packets cross the rings twice. The chaining application showcases the worst-case throughput since the \click elements are not performing any computation on the network packets. 

\begin{figure}[t]
 \centering
 \includegraphics[width=\columnwidth]{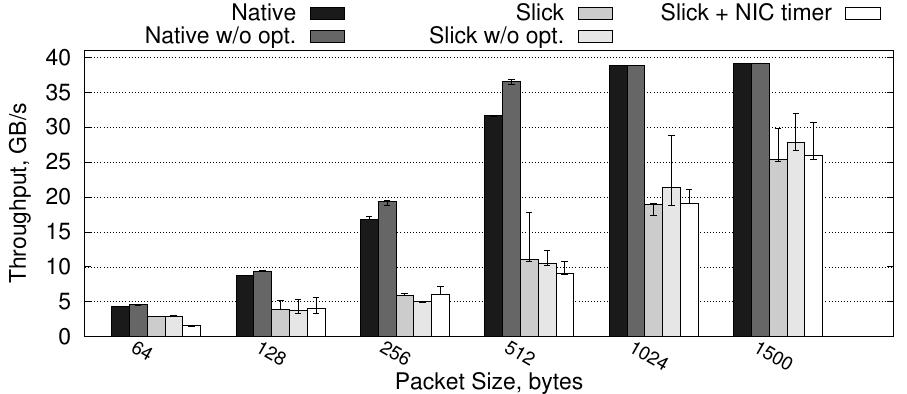}
 \caption{NFV chaining application throughput}
 \label{fig:complex-chaining}
\end{figure}

Figure~\ref{fig:complex-chaining} presents the throughput with varying packet size for the NFV chaining application. The results show that using the ring communication causes a substantial performance drop for \slick independent of the optimizations. This is mostly related to the way \scone runs enclaves---it must allocate constantly-running thread for the service threads created by \slick and \dpdk. Due to this, there is interference between the service threads and processing cores, which decreases the throughput and also increases the result variance.

\subsection{Packet Sealing Performance}
\label{subsec:eval:stateful}
\begin{figure}[t]
 \centering
 \includegraphics[width=\columnwidth]{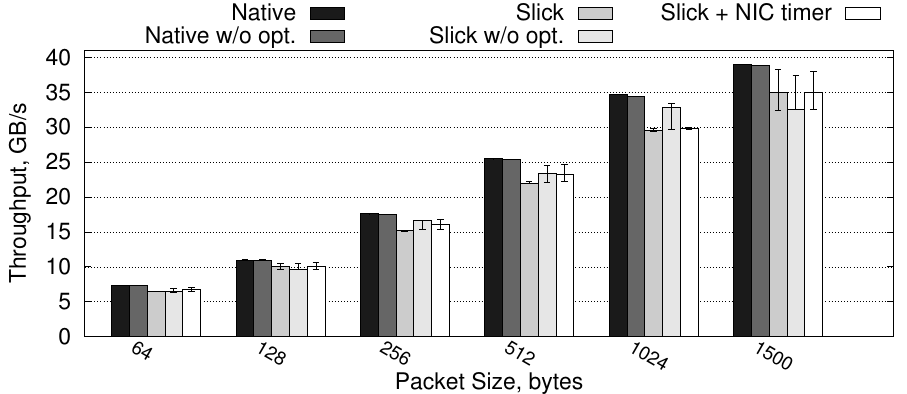}
 \caption{Seal application throughput}
 \label{fig:throughput-seal}
\end{figure}

We next evaluate the throughput of the {\tt Seal/Unseal} secure elements. In particular, we use our AES-GCM encryption code running inside the \sgx enclave (the same code is used for middlebox state persistence code). Figure~\ref{fig:throughput-seal} presents the throughput of the {\tt Seal} element with varying packet size. The result shows that the code inside \sgx enclave runs within $88$\% of the native performance irrespective of the optimizations applied. This is explained by the fact that most of the application CPU time is spent doing the encryption. The difference between the native and \sgx version can be explained by different thread scheduling strategies used by \scone and native {\tt POSIX}. In {\tt POSIX}, threads are pinned to the real CPU cores, while in \scone, the userspace threads inside enclave are pinned to the in-enclave kernel threads. This makes thread pinning non-deterministic---sometimes two threads that are to be pinned to different cores are pinned to sibling hyper-threads. 

\subsection{Case Studies}
\label{sec:eval-case-studies}

We next evaluate \sys's performance with the following two case-studies: (1) {\tt IPRouter}, and (2) {\tt IDS}.

\begin{figure}[t]
 \centering
 \includegraphics[width=\columnwidth]{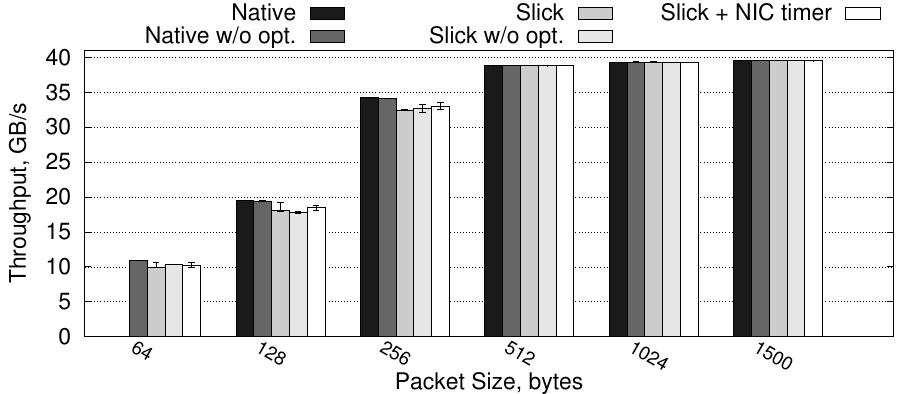}
 \caption{IPRouter throughput measurements}
 \label{fig:complex-router-tput}
\end{figure}

\myparagraph{IPRouter} {\tt IPRouter} application is an adaptation of a multi-port router \click example application to our evaluation hardware. This application first classifies all packets into three categories: ARP requests, ARP replies, and all other packets. ARP requests are answered. ARP replies are dropped. Other packets are passed to a routing table element that sends them to the NIC output port. Figure~\ref{fig:complex-router-tput} shows the throughput of the {\tt IPRouter} application with varying packet size.  We can see that \sys has the same performance as \click with packet sizes bigger than 256 bytes, and performs within 90\% of \click with smaller packets.

\begin{figure}[t]
 \centering
 \includegraphics[width=\columnwidth]{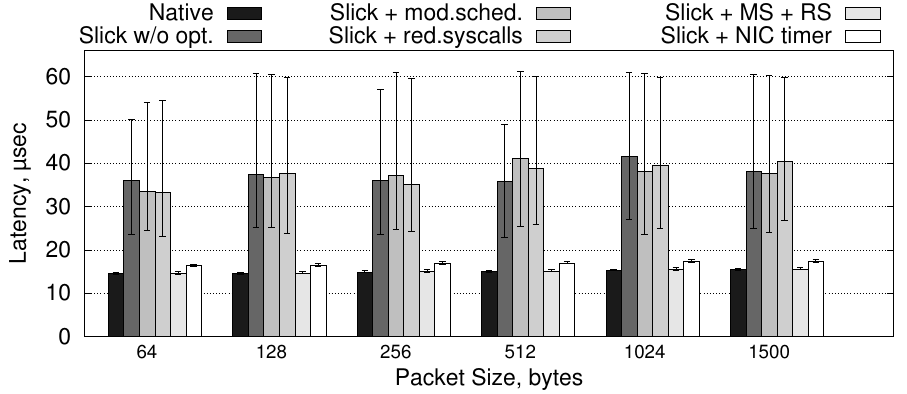}
 \caption{IPRouter latency measurements}
 \label{fig:latency-iprouter}
\end{figure}

We also measured the latency of the {\tt IPRouter} application as presented in Figure~\ref{fig:latency-iprouter}. We can see that even if the number of elements in the application increases, latency of the application remains the same as the native execution.

\myparagraph{Intrusion Detection System (IDS)} {\tt IDS} application implements NF that is commonly found in the enterprise network. It pushes the traffic through the firewall, and then performs traffic scanning with the {\tt HyperScan} element. Traffic that does not match any pattern is sent to the output, while matching traffic passes through a counter and then dropped.

\sys performs as close to the native \click execution with a slight performance drop. This drop comes from the general \sgx overhead for memory accesses. Due to the space constraints, we omit the latency measurement results for IDS. 

\begin{figure}[t]
 \centering
 \includegraphics[width=\columnwidth]{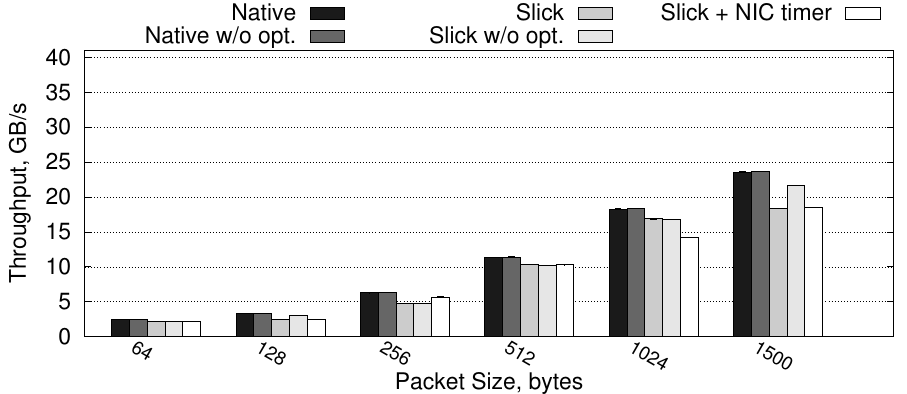}
 \caption{IDS throughput measurements}
 \label{fig:complex-ids}
\end{figure}

\section{Related Work}
\label{sec:related}

\myparagraph{Software middleboxes} \click's~\cite{click} modular architecture has been leveraged in the research community to build many useful software-based middlebox platforms~\cite{clickos, slick-sosr, openbox,clickNP, stateless-click, stateless-click}. Our work also builds on the \click architecture, but unlike the previous work, \slick focuses on securing the \click architecture on the untrusted commodity hardware. 

However, most \click-based middleboxes operate at L2-L3 layer, with the notable exception of CliMB~\cite{climb}. To support flow-based abstractions, many state-of-the-art  middleboxes~\cite{mos, comb, xomb, netagg, flick, netagg} support comprehensive applications and use-cases. Since both \click and \dpdk are geared toward L2-L3 network processing, our current architecture does not support L4-L7 network functions. As part of the future work, we plan to integrate a high-performance user-level networking stack~\cite{mTCP} in the \scone framework to support the development of secure higher layer middleboxes.
%
%
%
%
%
%

\myparagraph{Secure middleboxes} APLOMB~\cite{aplomb} is one the first systems to showcase that it is a viable alternative, performance- and cost-wise, to outsource middleboxes from the enterprise environment to the cloud. However, APLOMB does not consider the security implications of outsourcing in the cloud. 

To overcome the limitation of APLOMB, the follow up systems, namely Embark~\cite{embark} and BlindBox~\cite{blindbox}, advocate network data processing over the encrypted traffic. In particular, BlindBox~\cite{blindbox} proposes an encryption scheme based on garbled circuits to support string matching operations over encrypted traffic. However,  Blindbox supports only restrictive type of middlebox functionalities, such as NFs for deep packet inspection. To overcome this limitation, Embark~\cite{embark} extends BlindBox to support a wider range of middlebox functions. However, Embark suffers from prohibitively low performance as it  involves complex cryptographic computations over the encrypted network traffic. In contrast, \slick supports a wide range of network functions (same as plain-text), and achieves a near-native throughput and latency.


The recently published workshop papers~\cite{trusted-click, sgx-nfv} have elaborated the challenges and potential usages of \sgx in the network applications. In the domain of network-intensive applications, SGX-Tor~\cite{sgx-tor} is one of the first systems to use \sgx to enhance the security and privacy of Tor. In a similar vein, CBR~\cite{secure-cbr} leverages \sgx to support privacy-preserving routing. Likewise, the \slick project builds the first comprehensive system using \intel \sgx to secure middleboxes.

%
%
%

\myparagraph{Shielded execution} Shielded execution  provides strong security guarantees for legacy applications running on untrusted platforms~\cite{haven, sgxbounds, scone, vc3, graphene-sgx, panoply}. Our work leverages shielded execution based on \intel \sgx. It is worth noting that unlike the prior usage of shielded execution for commonly used services like HTTP servers or KV stores, we need to adapt the shielded execution to process the network traffic at line rates. To achieve this, \slick is the first system that integrates a high-speed packet I/O framework~\cite{dpdk, netmap, packetshader} with shielded execution.


%



\section{Conclusion}
\label{sec:conclusion}

In this paper, we presented the design, implementation, and evaluation of \sys{}---a secure middlebox framework for deploying high-performance Network Functions (NFs) on untrusted commodity servers.  \sys exposes a generic interface based on \click to design and implement a wide-range of NFs  using its out-of-the box elements and C++ extensions.   To securely process data at line rate, \sys is the first system to integrate a high-performance I/O processing library (\intel \dpdk) with a shielded execution (\scone) framework based on \intel \sgx.  We have also added several new useful features, and optimizations for secure network processing. Our evaluation using a wide-range of NFs and case-studies show that \sys{} achieves near-native throughput and latency. 

\myparagraph{Acknowledgements} This project was funded by the European Union's Horizon 2020 program under grant agreements No. 645011 (SERECA), No. 690111 (SecureCloud), and No. 690588 (SELIS).


%

\balance
\bibliographystyle{abbrv}
\bibliography{main} 

\end{document}